\documentclass[journal]{IEEEtran}
\ifCLASSINFOpdf
\usepackage[pdftex]{graphicx}
\graphicspath{figures/}
\DeclareGraphicsExtensions{.pdf,.jpeg,.png}
\else
\usepackage[dvips]{graphicx}
\graphicspath{{figures/}}
\DeclareGraphicsExtensions{.eps}
\fi
\usepackage[cmex10]{amsmath}
\usepackage{bm}

\usepackage{url}
\begin{document}
\title{Proposal for an All-Spin Artificial Neural Network: Emulating Neural and Synaptic Functionalities Through Domain Wall Motion in Ferromagnets}

\author{Abhronil~Sengupta,~\IEEEmembership{Student Member,~IEEE,}
Yong~Shim, 
and~Kaushik~Roy,~\IEEEmembership{Fellow,~IEEE}
\thanks{Manuscript received 13th October, 2015; revised 31st December, 2015; accepted 25th January, 2016. The work was supported in part by, Center for Spintronic Materials, Interfaces, and Novel Architectures (C-SPIN), a MARCO and DARPA sponsored StarNet center, by the Semiconductor Research Corporation, the National Science Foundation, Intel Corporation and by the National Security Science and Engineering Faculty Fellowship.}
\thanks{The authors are with the School
of Electrical and Computer Engineering, Purdue University, West Lafayette,
IN, 47907 USA. E-mail: asengup@purdue.edu.}}
\maketitle
\begin{abstract}
\small{Non-Boolean computing based on emerging post-CMOS technologies can potentially pave the way for low-power neural computing platforms. However, existing work on such emerging neuromorphic architectures have either focused on solely mimicking the neuron, or the synapse functionality. While memristive devices have been proposed to emulate biological synapses, spintronic devices have proved to be efficient at performing the thresholding operation of the neuron at ultra-low currents. In this work, we propose an All-Spin Artificial Neural Network where a single spintronic device acts as the basic building block of the system. The device offers a direct mapping to synapse and neuron functionalities in the brain while inter-layer network communication is accomplished via CMOS transistors. To the best of our knowledge, this is the first demonstration of a neural architecture where a single nanoelectronic device is able to mimic both neurons and synapses. The ultra-low voltage operation of low resistance magneto-metallic neurons enables the low-voltage operation of the array of spintronic synapses, thereby leading to ultra-low power neural architectures. Device-level simulations, calibrated to experimental results, was used to drive the circuit and system level simulations of the neural network for a standard pattern recognition problem. Simulation studies indicate energy savings by $\sim 100\times$ in comparison to a corresponding digital/ analog CMOS neuron implementation.}
\end{abstract}

\begin{IEEEkeywords}
Neuromorphic Computing, Artificial Neural Networks, Spintronics, Spin-Orbit Torque, Domain Wall.
\end{IEEEkeywords}

\section{Introduction}
The basic computing element in an Artificial Neural Network (ANN) involves weighted summation of neuron inputs followed by a thresholding operation. For instance, if the neuron's inputs are represented by $I_{i}$ and the corresponding synaptic weights are represented by $w_{i}$, then the neuron's output is represented by, $y=f(\sum w_{i}.I_{i}$). Here, $f$ represents the transfer function of the neuron, which could be a step, linear or sigmoid function of the resultant neuron input. Complex deep learning architectures used for pattern recognition tasks consists of interconnected layers of these basic computing blocks \cite{schmidhuber2015deep}. However, implementation of such algorithms on general-purpose computers are extremely area and energy inefficient since the sequential von-Neumann computing model is a complete contrast to the parallel, event-driven processing in the brain. Even custom analog/digital CMOS implementations have proved to be several orders of magnitude higher power and area consuming since they do not offer a direct mapping to the weighted summation and thresholding operations involved in neural computation. While analog designs are power-hungry, digital designs tend to be area expensive.

As a result, significant research efforts are being directed to develop neural architectures based on emerging post-CMOS technologies like Phase Change Memories (PCM) \cite{jackson2013nanoscale,kuzum2011nanoelectronic}, Ag-Si memristors \cite{jo2010nanoscale}, etc. Inspired by the fact that synapses are the main seat of learning and that synapses outnumber the neurons by several orders of magnitude, researchers have demonstrated how such nanoelectronic devices can directly mimic synapse functionalities \cite{jackson2013nanoscale,kuzum2011nanoelectronic,jo2010nanoscale,rajendran2013specifications,ramakrishnan2011floating,sengupta2015spins}. However, since these synapses are required to drive analog CMOS neurons, power consumption of these neuromorphic systems are still orders of magnitude higher in comparison to that involved in the brain \cite{sharad2013spin}.

Recent discoveries in spintronics have brought forward a set of physical phenomena that offer a direct mapping to the thresholding operation of neurons. Researchers have proposed Lateral Spin Valve (LSV) structures \cite{sharad2012spin}, domain wall based devices \cite{sharad2013spin} and spin-Hall effect based structures \cite{:/content/aip/journal/apl/106/14/10.1063/1.4917011} that can be operated at ultra-low voltages and can be switched by ultra-low currents. Resistive crossbar array of programmable synapses based on memristors or PCMs can be interfaced with these spintronic neurons to perform ultra-low power non-Boolean computation \cite{sharad2013spin}. However, a basic limitation of such computing platforms was that these spin-neurons could only emulate the step transfer function in ANNs (corresponding to the switching of a nanomagnet between its two stable states by an input synaptic current) whereas, non-step (linear or sigmoid) transfer function is more attractive for complex pattern recognition tasks since more information can be encoded in the neuron's output in the latter case. 
\begin{figure*}[!t]
\centering
\includegraphics[width=6.5in]{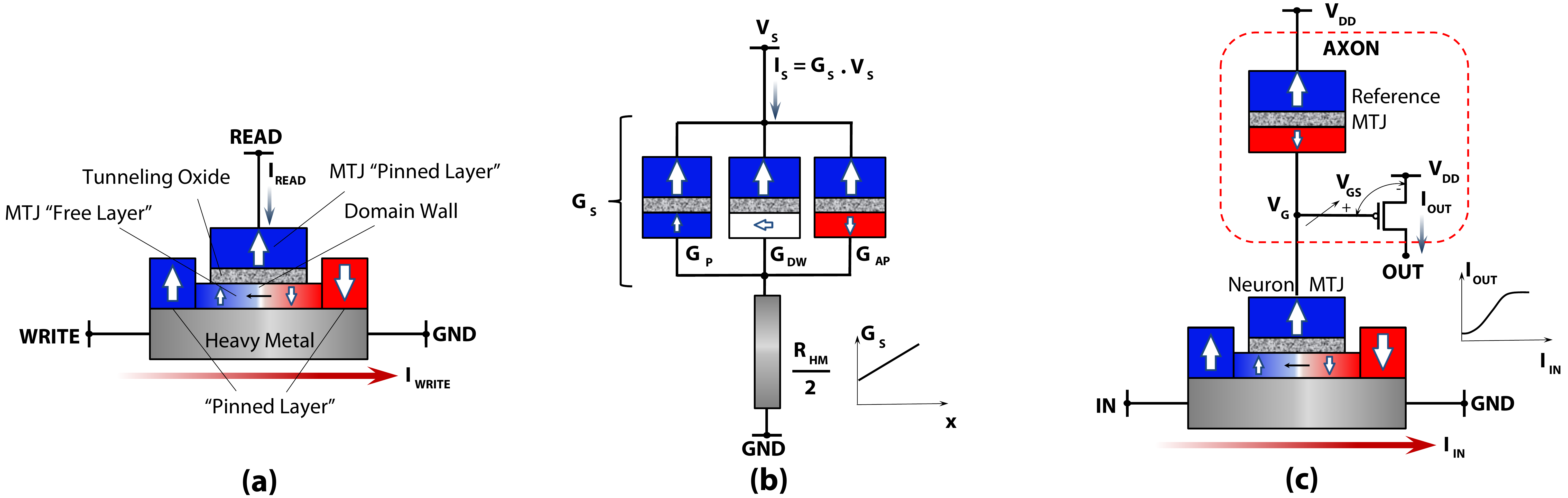}
\caption{(a) Three terminal device structure acting as the basic building block for the All-Spin ANN. Spin-orbit torque (SOT) generated by current, $I_{write}$, through the heavy metal programs the domain wall position in the MTJ ``free layer". The domain wall position encodes the device conductance between terminals READ and GND. (b) Operation of the spintronic device as a synapse. The ``read" current flowing through the device gets modulated by the MTJ conductance. The conductance encodes the synaptic weight and can be set by programming the domain wall position. (c) Operation of the spintronic device as a neuron. Initially the neuron is ``reset'' such that the domain wall position is initialized to the left edge of the ``free layer''. Then the resultant synaptic input current programs the domain wall position. Subsequently, during the ``read'' phase, the ``Reference MTJ" and PMOS transistor serve as the axon to propagate the neuron output to its fan-out neurons. The transfer function of the neuron is characterized by the relationship between $I_{OUT}$ and $I_{IN}$. }
\label{fig1}
\end{figure*}
In this paper, we explore the possibility of an All-Spin neuromorphic architecture where the core element is a spintronic device based on a ferromagnet (FM)-heavy metal (HM) multilayer structure, to realize an ultra-low power neural computing platform. The ferromagnet consists of a domain wall separating two oppositely polarized magnetic domains and is programmed by spin-orbit torque (SOT) generated by the heavy metal underlayer. We demonstrate the mapping of this single device to neuron and synapse functionalities and illustrate how spintronic synapses interfaced with spintronic neurons can communicate via CMOS axon transistors to form a neural network. Further, the device can provide a non-step transfer function and thereby has the potential to be utilized for complex pattern recognition problems. The paper is organized into the following sections. Section II provides a brief description of the underlying device physics in the spintronic device. Section III illustrates the functionality of the proposed device as a neuron and a synapse. Section IV explains the All-Spin neuromorphic architecture in details. Section V describes the simulation framework along with simulation results. 
\section{Spintronic Device Structure: Principle of Operation} 
Let us first provide a brief discussion on the underlying device physics and principle of operation of the three terminal device structure (Fig.\ref{fig1}a) that serves as the basic building block for the All-Spin ANN. The device is a Magnetic Tunnel Junction (MTJ) where the ``free layer" (ferromagnet whose magnetization can be manipulated) is separated from the ``pinned layer" (ferromagnet whose magnetization is fixed) by a tunneling oxide barrier (MgO). The ``free layer" consists of a ferromagnet where a domain wall separates two oppositely polarized magnetic regions. 

\begin{figure}[!b]
\centering
\includegraphics[width=3.0in]{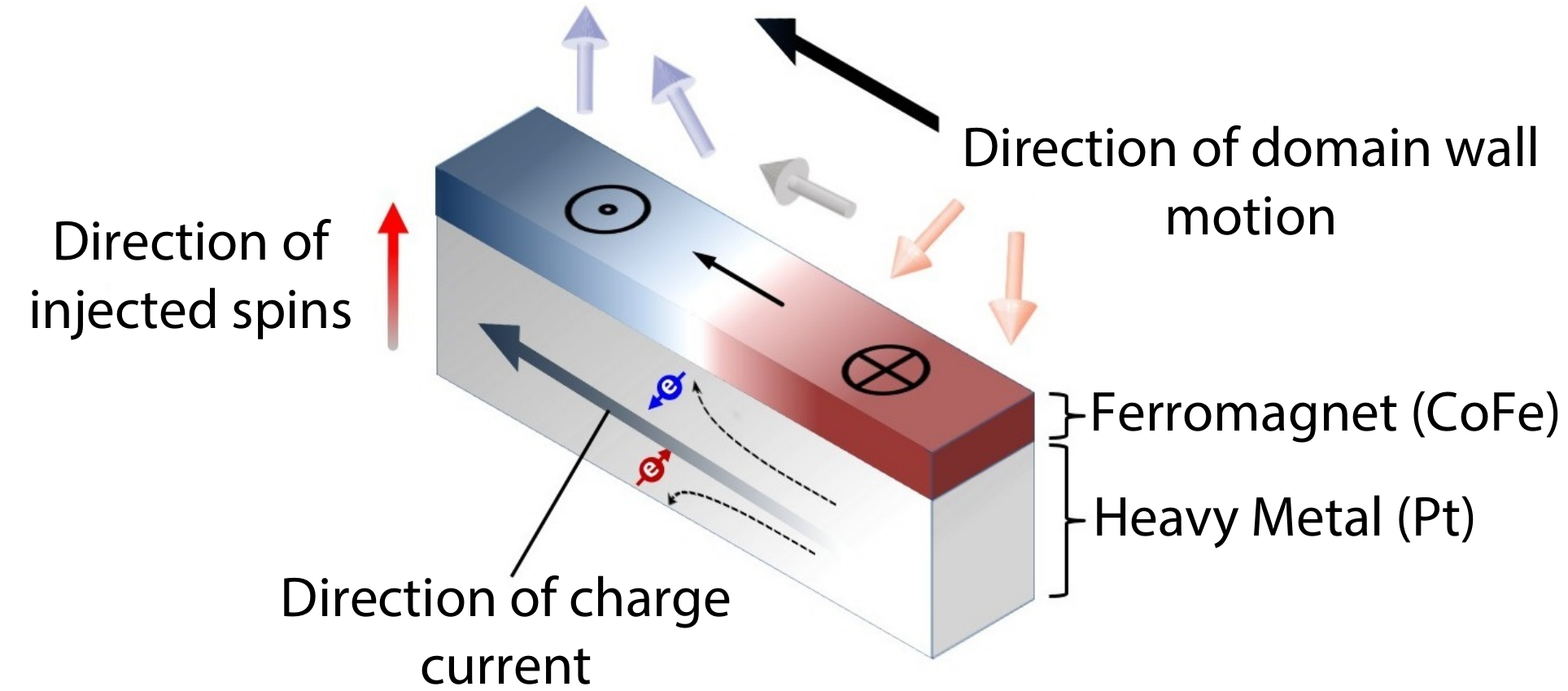}
\caption{Underlying physical phenomena responsible for domain wall motion in PMA nanowires due to the flow of an in-plane charge current through the HM underlayer. Spin-orbit coupling causes stabilization of chiral N\'{e}el domain wall in a ferromagnet-heavy metal heterostructure through the Dzyaloshinskii-Moriya Interaction (DMI). Due to spin-Hall effect, a transverse spin current is produced by the in-plane charge current flowing through the HM which causes movement of the domain wall.}
\label{physics}
\end{figure}

The domain wall in the ferromagnet (FM) can be displaced by SOT exerted by a charge current flowing through a heavy metal (HM) underlayer. Current flow through an underlying HM in FM-HM heterostructures has recently become a promising mechanism to achieve deterministic domain wall displacement in FMs ~\cite{emori2013current,martinez2014current,emori2014spin,ryu2014chiral,ryu2013chiral}. This is mainly due to the fact that the same domain wall displacement can be achieved by current density magnitudes that are $\sim 100\times$ lower in comparison to conventional spin-transfer torque driven domain wall motion. Further, the input charge current flows mainly through the HM underlayer whose resistance is generally an order of magnitude lower than that of the FM. It is worth noting here that no external magnetic field is required for the displacement of the domain wall. Spin-orbit coupling at the interface of such magnetic multilayers (with Perpendicular Magnetic Anisotropy) leads to Dzyaloshinskii-Moriya exchange interaction (DMI) which results in the stabilization of a chiral N\'{e}el domain wall~\cite{emori2013current,martinez2014current,emori2014spin,ryu2014chiral,ryu2013chiral}. Assuming spin-Hall effect ~\cite{emori2013current,martinez2014current,emori2014spin,ryu2014chiral,ryu2013chiral} to be the dominant underlying physical phenomena, an in-plane charge current through the HM underlayer results in the generation of a transverse spin current due to deflection of opposite spin polarizations at the top and bottom surfaces of the HM. For magnetic multilayers with left-handed chirality (see Fig.\ref{physics}), input charge current flow through the HM underlayer results in domain wall movement in the same direction and vice-versa. 

Fig.\ref{physics} illustrates the underlying physical phenomena responsible for domain wall motion in magnetic heterostructures with perpendicular magnetic anisotropy (PMA) due to the flow of an in-plane charge current through a heavy metal underlayer. The magnetization dynamics of the ferromagnet can be described by solving Landau-Lifshitz-Gilbert equation with additional term to account for the spin-orbit torque generated by spin-Hall effect (SHE) at the FM-HM interface ~\cite{martinez2014current,slonczewski1989conductance},
\begin{equation}
\frac {d\widehat {\textbf {m}}} {dt} = -\gamma(\widehat {\textbf {m}} \times \textbf {H}_{eff})+ \alpha (\widehat {\textbf {m}} \times \frac {d\widehat {\textbf {m}}} {dt})+\beta (\widehat {\textbf {m}} \times \widehat {\textbf {m}}_P \times \widehat {\textbf {m}})
\end{equation}
where $\widehat {\textbf {m}}$ is the unit vector of FM magnetization at each grid point, $\gamma= \frac {2 \mu _B \mu_0} {\hbar}$ is the gyromagnetic ratio for electron, $\alpha$ is Gilbert\textquoteright s damping ratio, $\textbf{H}_{eff}$ is the effective magnetic field, $\beta=\frac{\hbar \theta J}{2 \mu_0 e t M_s}$ ( $\hbar$ is Planck’s constant, $J$ is input charge current density, $\theta$ is spin-Hall angle~\cite{martinez2014current}, $\mu_0$ is permeability of vacuum, $e$ is electronic charge, $t$ is FL thickness and $M_s$ is saturation magnetization) and $\widehat {\textbf {m}}_P$ is direction of input spin current. The effective field $\textbf{H}_{eff}$ also includes the field due to DMI~\cite{martinez2014current} and is given by,
\begin{equation}
\textbf{H}_{DMI} = -\frac{2D}{\mu_{0}M_{s}}\left[\frac{\partial m_{z}}{\partial x}\widehat{x} + \frac{\partial m_{z}}{\partial y}\widehat{y} - \left(\frac{\partial m_{x}}{\partial x}+\frac{\partial m_{y}}{\partial y}\right )\widehat{z}\right ]
\end{equation}
Here, $D$ represents the effective DMI constant and determines the strength of DMI field in such multilayer structures. A positive sign of $D$ implies right-handed chirality and vice versa. In the presence of DMI, the boundary conditions at the edges of the sample is given by,
\begin{equation}
\frac{\partial \widehat {\textbf {m}}}{\partial n} = \frac{D}{2A}\widehat {\textbf {m}} \times \left(\widehat {\textbf {n}} \times \widehat{z} \right )
\end{equation}
where, $A$ is the exchange correlation constant and $\widehat {\textbf {n}}$ represents the unit vector normal to the surface of the FM. Current density was estimated by assuming that the current flow is mainly through the FM-HM layers in the stack structure~\cite{martinez2014current}.

Based on this physical phenomena, we propose a three terminal device structure, as shown in Fig.\ref{fig1}a, with decoupled ``write" and ``read" current paths. The ``write" current, $I_{write}$, flows through the HM underlayer (between terminals WRITE and GND) and its magnitude determines the position of the domain wall in the MTJ ``free layer". The ``pinned layer"s on either side of the ``free layer" serve to stabilize the domain wall at either side of the ``free layer" for large magnitudes of the input current. On the other hand, the ``read" current, $I_{read}$, flows through the MTJ structure (between terminals READ and GND). The magnitude of the ``read" current is modulated by the device conductance, which in turn, is a function of the domain wall position in the ``free layer" of the MTJ. The domain wall position can be programmed by an appropriate charge current, $I_{write}$, between terminals WRITE and GND. The ``read" current magnitude has to be maintained lower than the minimum current responsible for domain wall depinning. It is worth mentioning here, that deterministic domain wall movement have been experimentally demonstrated in such magnetic multilayer structures ~\cite{emori2013current,martinez2014current,emori2014spin,ryu2014chiral,ryu2013chiral}. However, MTJs have been traditionally used to ``read'' the ``free layer'' magnetization state at the two extreme resistance states, namely the parallel (P) or the anti-parallel (AP) state. In this proposal, we exploit the analog resistance variation of the MTJ with change in domain wall position to realize neuron and synapse functionalities in such spintronic devices. We believe that this proposal for an All-Spin ANN will stimulate proof-of-concept experiments to develop such non-Von Neumann device structures suitable for low-power neural processing. 

In order to simulate the variation of the MTJ resistance with applied voltage and oxide thickness, Non-Equilibrium Green's Function (NEGF) based transport simulation framework \cite{fong2011knack} was utilized. Considering that the FM has a uniform magnetization direction, the MTJ resistance ($R$) is a function of the spacer (MgO) thickness ($t_{MgO}$), relative angle between the magnetizations of the FM and the pinned layer ($\theta$), and the voltage across the MTJ ($V_{MTJ}$). The variation is described by the following equations,
\begin{equation}
R \propto \left( e^{a_{0}t_{MgO}+b_0} + \sum\limits_{m=1}^c \left( (-1)^{m-1}V_{MTJ}^{2m}e^{a_m t_{MgO}+b_m}\right) \right)^{-d}
\end{equation}

\begin{equation}
R(\theta) = \left( \frac{1}{R_P} \left( \cos\left(\frac{\theta}{2}\right)\right)^2 + \frac{1}{R_{AP}} \left( \sin\left(\frac{\theta}{2}\right)\right)^2 \right)^{-1}
\end{equation}
Here, $R_{P}$ and $R_{AP}$ represent the parallel ($\theta =0$) and anti-parallel resistance ($\theta =\pi$) of the MTJ respectively. The fitting parameters $a_m, b_m, c$ and $d$ are determined by calibrating the simulation framework with experimental data reported in \cite{yuasa2004giant,lin200945nm}. For an extensive description of the NEGF based simulation framework, readers are referred to Ref. \cite{fong2011knack}.
\begin{figure}[!t]
\centering
\includegraphics[width=3.3in]{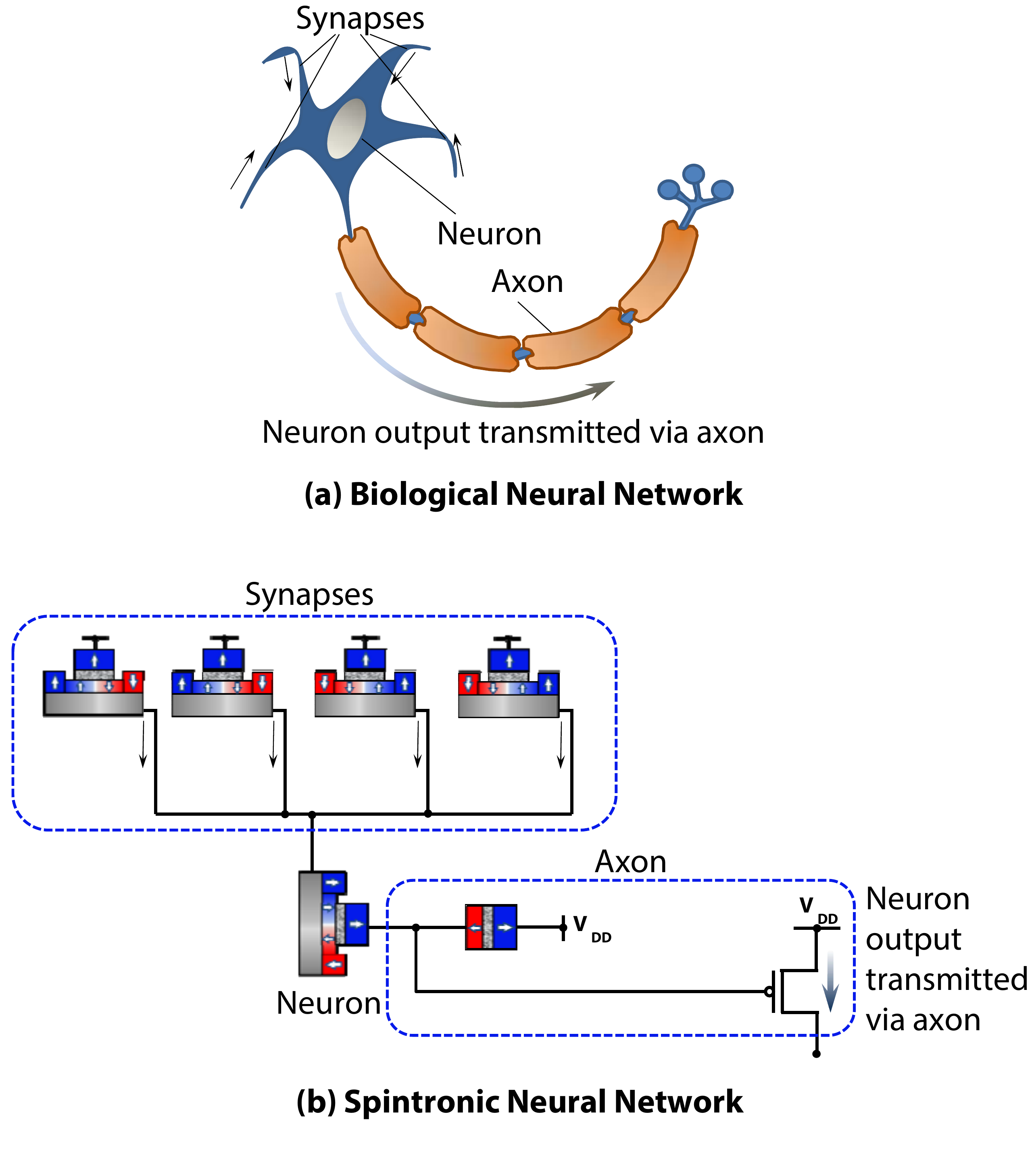}
\caption{(a) Biological neuron receives synaptic inputs and generates a corresponding output which is then transmitted via the axon. (b) All-Spin ANN where spintronic devices directly mimic neuron and synapse functionalities and axon (CMOS transistor) transmits the neuron's output to the next stage. During the ``neuron'''s write stage, the synapses will provide a resultant input current which is the weighted summation of the voltages applied across them. Note that voltage drop across magneto-metallic spin-``neuron''s is extremely small. During the ``read'' stage, the ``neuron'' transmits its output to the next fan-out stage via the axon transistor.}
\label{fig2}
\end{figure}

\section{Neuron and Synapse Functionalities of the Proposed Device}

\subsection{Synapse Functionality}
The synapse functions as the memory element in neuromorphic architectures. Synapses are junctions (characterized by weights) between neurons and transmit weighted signals from the transmitting to the receiving neuron. The operation of the spintronic device as a synapse is explained in Fig.\ref{fig1}b. For a fixed applied voltage at the READ terminal, the resistance of the device between the READ and GND terminals is mainly given by the parallel combination of the AP and P domains along with a small region where the ``free layer" magnetization is along the horizontal axis (corresponding to the domain wall region). The heavy-metal resistance in the path of $I_{read}$ is negligible in comparison to the tunneling oxide resistance. Let us denote the device conductance when the entire ``free layer" magnetization is P (AP) to the ``pinned layer" by $G_{P,max} (G_{AP,max})$. Thus, for an intermediate position $x$ of the domain wall from the left-edge of the MTJ, the equivalent conductance will be given by,
\begin{equation}
G_{S} (x) = G_{P,max}\left(\frac{x}{L}\right)+G_{AP,max}\left(\frac{1-x}{L}\right)+G_{DW}.
\end{equation}
Here, $G_{DW}$ denotes the conductance of the domain wall region and $L$ represents the length of the MTJ (excluding the domain wall width). For a fixed voltage applied between READ and GND terminals, $G_{P,max}$, $G_{AP,max}$ and $G_{DW}$ are constants. Hence, $G_{S}$ is a linear function of $x$ and therefore, the conductance in the path of the read current can be appropriately set by programming the domain wall position (by passing ``write" current through the heavy metal). Hence, when a voltage $V_{S}$ is applied across the spintronic synapse, the current, $I_{S}=G_{S}.V_{S}$, flowing through the device gets modulated by the MTJ equivalent conductance (which encodes the synaptic weight). It is worth noting here, that the resistance range of the spintronic synapses can be varied by varying the oxide thickness. Additionally, the critical current responsible for domain wall depinning for a particular time duration scales linearly with the device width. Hence, the width of the spintronic synapses can be appropriately designed such that the ``read" current through the MTJ (which flows through some portion of the heavy metal) does not cause any domain wall motion. Although there is some injected spin current due to ``read'' current flowing through the PL, its magnitude is much smaller in comparison to that due to SOT.

The ratio of the maximum to minimum synaptic weight encoded in the MTJ conductances will be determined by the Tunneling Magnetoresistance Ratio (TMR) values of the devices. MTJ TMR values of $\sim 600\%$ \cite{ikeda2008tunnel} have been fabricated resulting in a maximum to minimum weight ratio of $\sim 7\times$. TMR values $> 1,000\%$ are expected within a time period of ten years \cite{hirohata2015roadmap}.
\subsection{Neuron Functionality}
The neuron serves as the computing element in ANNs. It is characterized by a transfer function, i.e. it produces an output signal in accordance to the magnitude of its resultant synaptic input signal. The neuron's output signal is transmitted via the axon as an input to its fan-out neurons. Fig.\ref{fig1}c demonstrates the operation of the proposed spintronic device as a neuron. The neuron operation takes place in three stages, namely the ``write", ``read" and ``reset" stages. During the ``write" stage, the neuron (denoted by ``Neuron MTJ") receives the resultant synaptic input current (flowing between terminals IN and GND) and its magnitude determines the domain wall position. Higher the magnitude of the synapse current, higher is the domain wall displacement and hence lower is the device resistance. During the ``read" cycle, the IN terminal is deactivated and the ``axon" circuit is activated. The ``Reference MTJ" (whose orientation is fixed in the AP state) serves to produce a resistive divider network such that the gate voltage $V_{G}$ of the PMOS transistor decreases (due to decrease in resistance of the pull-down network) with increase in the magnitude of the synaptic input current. Hence, the output current, $I_{OUT}$, provided by the PMOS transistor increases with increase in the magnitude of the input current, $I_{IN}$ (due to increase in magnitude of $V_{GS}$ of the PMOS transistor). The neuron transfer function is characterized by the relationship between $I_{OUT}$ and $I_{IN}$ and the output transistor mimics the axon functionality of biological neurons by propagating the neuron output signal to fan-out neurons in the next stage. Finally, during the ``reset" phase the domain wall is initialized to the left edge of the ``free layer" of the MTJ neuron for the next operation cycle.

\subsection{Correspondence to Biological Neural Network}
Fig.\ref{fig2} demonstrates the close correspondence between the biological neural network and the proposed All-Spin neural network and thereby illustrates the direct mapping of synapse and neuron functionalities to nanoelectronic spin-devices. The biological neuron receives synaptic inputs from synapses and generates an output that is a function of the resultant input. The neuron's output is transmitted via the axon to fan-out neurons. Similarly, the spintronic neuron receives a resultant synaptic current which is the weighted summation of its inputs. This resultant current input flowing through the heavy metal of the spintronic neuron generates an output which is transmitted via the CMOS transistor, acting as the axon, to the next stage. 
\section{All-Spin Neuromorphic Architecture}
\begin{figure}[!t]
\centering
\includegraphics[width=3.0in]{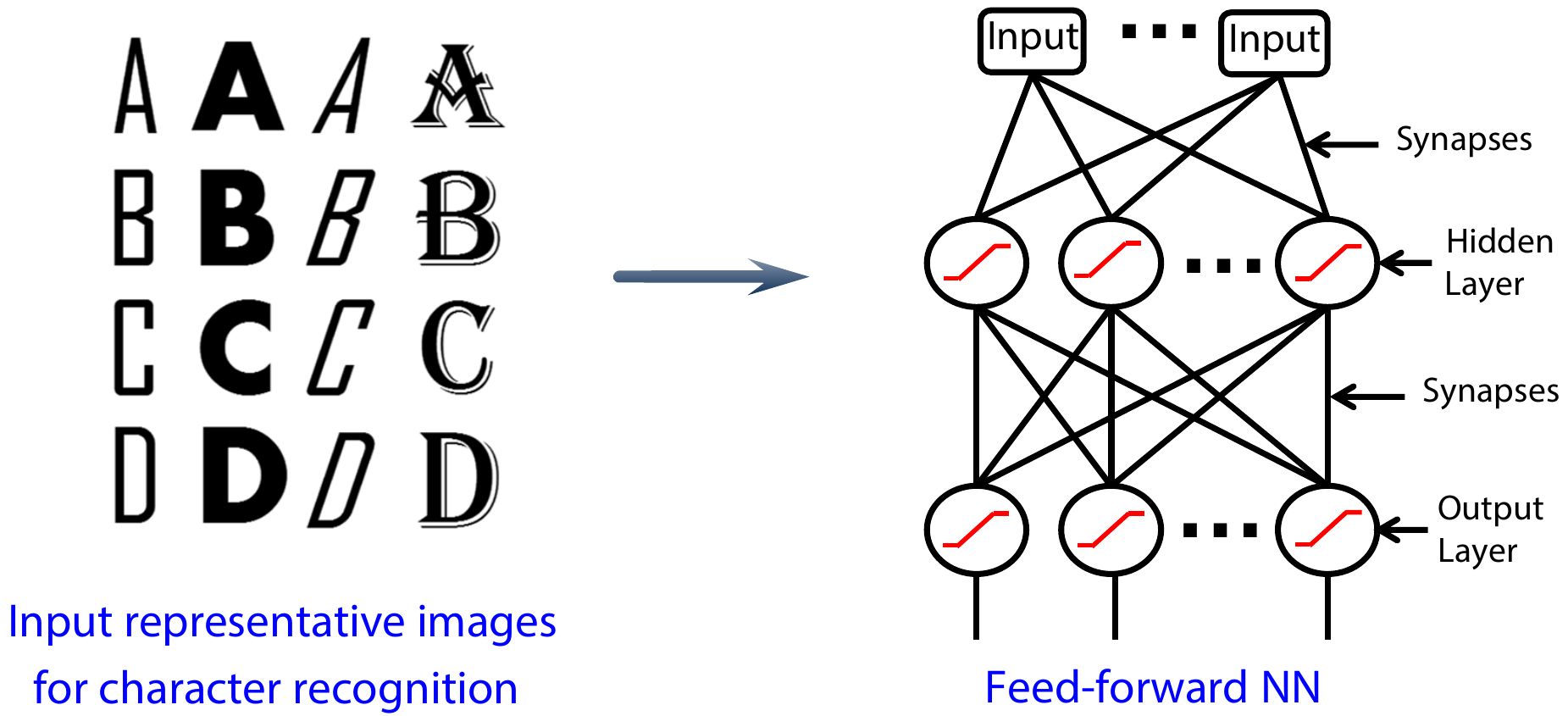}
\caption{Input representative images of characters that were used for pattern recognition and the feedforward neural network comprising of a hidden layer and an output layer that was utilized for the purpose.}
\label{chars74}
\end{figure}
\begin{figure}[!t]
\centering
\includegraphics[width=3.3in]{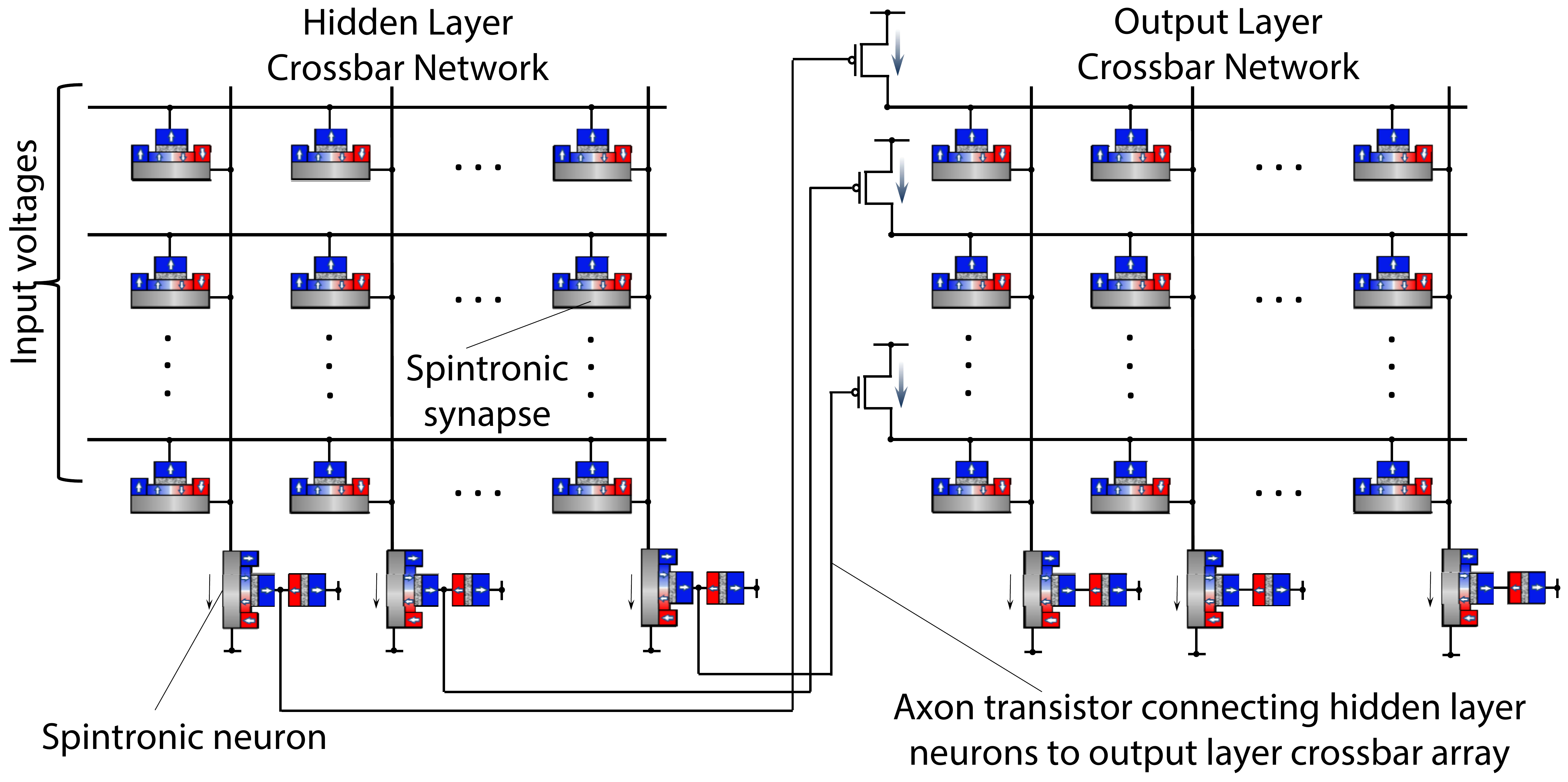}
\caption{Hardware mapping of the ANN into an All-Spin neural architecture. Neurons in the hidden layer receives inputs through weighted synapses and transmits their output to each neuron in the output layer through synapses. The spin devices present at each cross-point encode the synaptic weight and provide a resultant synaptic current to the spin-neuron. Axon transistors interconnect the hidden layer and the output layer crossbar arrays. More number of such interconnected layers can be used to realize compact and energy-efficient deep learning architectures. }
\label{arch}
\end{figure}
Next, let us consider the hardware mapping of a feed-forward ANN comprising of a hidden layer of neurons connected in an all-to-all fashion to an output layer (Fig.\ref{chars74}). Let us first discuss the operation of the hidden layer. The hidden layer can be represented by a resistive crossbar network, as shown in Fig.\ref{arch}, where the input voltages (corresponding to pixel intensities) are applied along the horizontal rows. The vertical columns are connected to the spintronic neurons. Spintronic synapses are present at each cross-point of the array and the domain wall position in the device encodes the value of the corresponding synaptic weight or conductance. Let us denote the synaptic device conductance connecting the $i$-th input to the $j$-th neuron as $G_{ij}$ and the neuron resistance lying in the path of the resultant synaptic current as $R_j$. Equating the current supplied by the resistive synapses to the current flowing through the neuron, the net synaptic current supplied to the neuron is given by, 
\begin{equation}
I_j=\frac {\sum\limits_{i}G_{ij}.V_{i}} {1+ \gamma}.
\end{equation} 
where, $\gamma=R_{j}\sum\limits_{i}G_{ij}.$ If $\gamma<<1$, i.e. the neuron resistance is very small in comparison to 1/ $\sum\limits_{i}G_{ij}$, then the voltage drop across the spin-neurons can be practically ignored and consequently the resultant input current to the neuron will be a weighted summation of the voltage inputs. It is worth noting here, that the maximum input voltage to the crossbar array determines the range of synaptic resistances in order to ensure the critical current requirement for domain wall displacement between the two extreme edges of the neuron ``free layer". Higher is the input voltage, higher is the range of synaptic resistances and hence lower is the value of $\gamma$. However, in order to ensure that the synaptic resistance for a particular domain wall position does not vary significantly with applied voltage, the operating voltage has to be kept in the range of a few tens of $mV$. The low resistance magneto-metallic spintronic neurons operate at small terminal voltages less than $\sim 10mV$ and enable the ultra-low voltage operation of the crossbar array.

Depending on the resultant synaptic input current, the domain wall position gets programmed in the ``hidden layer" spintronic neurons. During the operation of the ``output layer", the ``read" circuit of the ``hidden" layer neurons is activated. The output axon PMOS transistor generates an output current which provides the input to the next layer. Considering the voltage drop across the spin-neurons for the next layer to be negligible, each PMOS transistor drives an equivalent conductance $G_{eq}$, which is the sum of the synaptic conductances for a particular row. The input voltage provided to each row of the ``output" layer crossbar array will be given by the product of the transistor output current and $G_{eq}$. In order to ensure that the input voltage does not vary with varying ${G}_{eq}$ for the different rows, a dummy column was considered in the crossbar array, where the conductance in a particular row of the dummy column is set such that the value of $G_{eq}$ is equal for all rows of the array. As explained earlier, the ``output" layer neurons receive currents proportional to the weighted summation of the inputs and produces a corresponding output.

It is worth mentioning here that the synaptic weights in an ANN can be positive or negative. In order to implement this functionality, the crossbar array for a particular layer can be split up into two separate ``positive" and ``negative" arrays. In case a particular synaptic weight is positive (negative), then the corresponding conductance in the ``positive" (``negative") array is set in accordance to the weight, else it is set to a very high OFF resistive state. The neuron ``write" operation is then split up into two cycles where the ``positive" crossbar array provides input current to the neuron in one direction during the first cycle and the ``negative" array provides input current in the opposite direction during the next cycle. 

Scalability of the proposed architecture can be determined by the driving capability of the neurons and synapses. Higher the number of synaptic inputs to the neuron, higher will be the magnitude of $\sum\limits_{i}G_{ij}$. However, the resistance range of the synapses can be appropriately tuned such that the ratio $\gamma<<1$. The number of neurons in the succeeding layer that can be connected to a particular neuron of the previous layer via the axon transistor is limited by the current driving capability, i.e. size of the transistor. However, this can be overcome by distributing large synaptic arrays into smaller crossbar arrays. 
\section{Simulation Framework and Results}
In order to assess the functionality and power consumption of the proposed All-Spin neural network, a hybrid device-circuit-algorithm co-simulation framework was used. The synergistic simulation framework, consisting of a ``top-down" and ``bottom-up" perspective, is described next.
\subsection{Top-Down Perspective}
As proof of concept, a small-scale ANN with 20 hidden layer neurons and 26 output layer neurons was used to recognize characters A-Z from the Chars74K dataset \cite{de2009character}. The input images were downscaled to size 16x16 and were applied as a 1-D vector to the input layer. The neuron transfer function was taken to be linearly increasing with the input, ultimately saturating at a maximum value. The choice of the transfer function was obtained from device and circuit level simulations and will be described later. Standard backpropagation algorithm was used to generate a set of weights and biases of the network (for mapping to synapse conductance values). The accuracy of the network over a set of 260 images from the dataset was evaluated to be $\sim 80\%$. 
\subsection{Bottom-Up Perspective}
\begin{figure}
\centering
\includegraphics[width = 3.3in ]{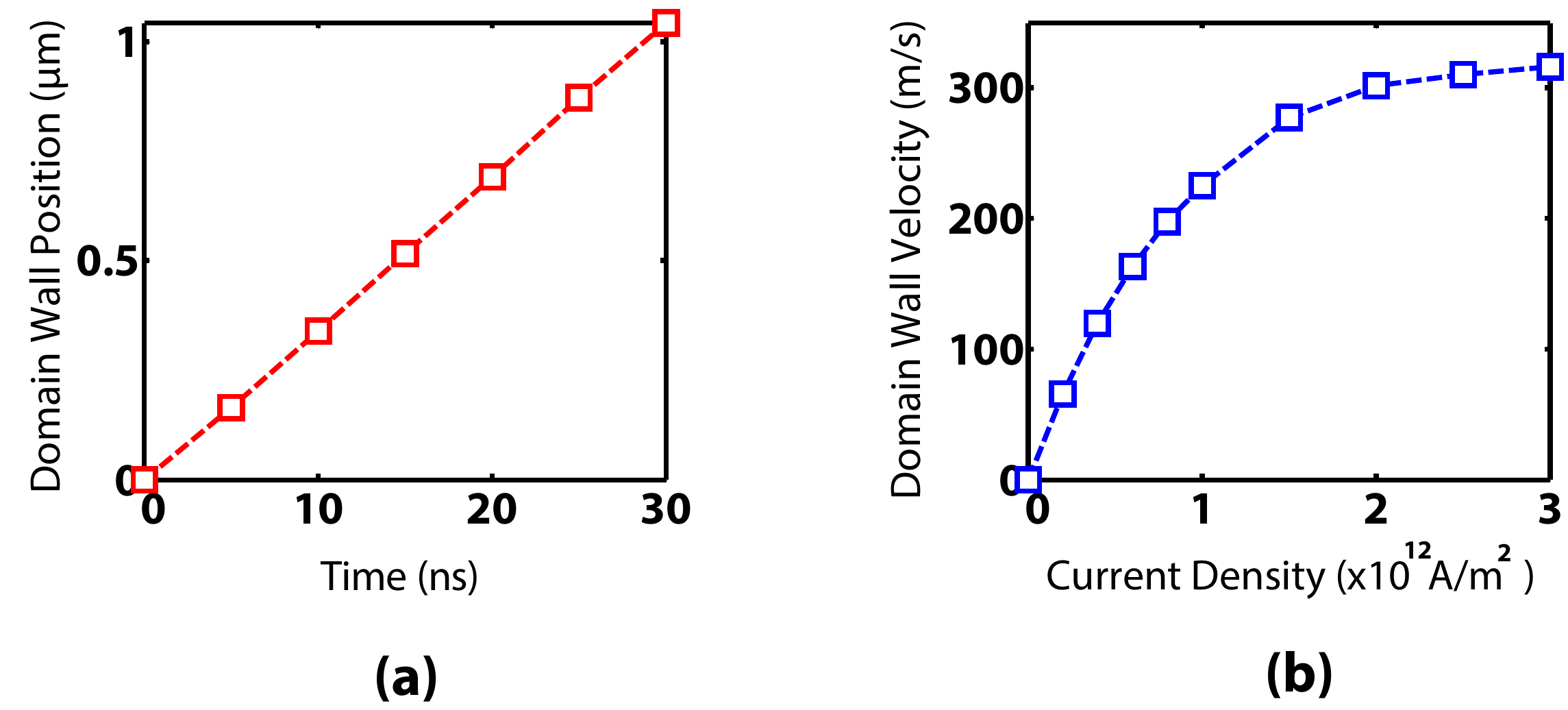}
\caption{(a) Domain wall displacement as a function of time for a CoFe strip of cross-section 160$nm \times$0.6$nm$ due to the application of a charge current density, $J= 0.1 \times 10^{12} A/m^2$, (b) Domain wall velocity as a function of current density. The domain wall displacement increases linearly with the magnitude of the charge current density and ultimately saturates to a maximum value. The simulation parameters (given in Table I) were obtained experimentally from magnetometric measurements of Ta(3nm)/Pt(3nm)/CoFe(0.6nm)/MgO(1.8nm)/Ta(2nm) nanostrips \cite{emori2013current,martinez2014current}.The graphs are in good agreement with \cite{martinez2014current}.}
\label{fig:calib}
\end{figure}
The bottom-up simulation framework involves the investigation of the device physics of current induced spin-orbit torque in ferromagnets and the development of behavioral models for system level simulations of the All-Spin neural network. The simulation parameters (given in Table I) were obtained experimentally from magnetometric measurements of Ta(3nm)/Pt(3nm)/CoFe(0.6nm)/MgO(1.8nm)/Ta(2nm) nanostrips \cite{emori2013current,martinez2014current}. Current density was estimated by assuming that the current flow is mainly through the ferromagnet-heavy metal layers in the stack structure \cite{emori2013current,martinez2014current}.

\begin{table}[h]
\renewcommand{\arraystretch}{1.3}
\caption{Simulation Parameters}
\label{table_1}
\centering
\begin{tabular}{c c}
\hline 
\bfseries { Parameters} & \bfseries { Value}\\
\hline
{ Ferromagnet Thickness} & { $0.6 nm$} \\
{ Grid Size} & { $ 4 \times 1 \times 0.6 nm^3$} \\
{ Heavy Metal Thickness} & { $ 3 nm$} \\
{ Domain Wall Width} & { $ 7.6 nm$} \\
{ Saturation Magnetization, $M_s$} & { 700 $KA/m$} \\
{ Spin-Hall Angle, $\theta$} & { 0.07} \\
{ Gilbert Damping Factor, $\alpha$} & { 0.3} \\
{ Exchange Correlation Constant, $A$} & { $1 \times 10^{-11} J/m$} \\
{ Perpendicular Magnetic Anisotropy, $K_{u2}$} & { $4.8 \times 10^{5} J/m^{3}$} \\
{ Effective DMI constant, $D$} & { $-1.2 \times 10^{-3} J/m^{2}$} \\
{ Resistivity of Pt, $\rho$} & { $200\Omega.nm$} \\
\hline
\end{tabular}
\end{table}

Fig. \ref{fig:calib}(a) shows the domain wall displacement in a CoFe sample with cross-section of $160nm \times 0.6nm$ for a charge current density of $J= 0.1 \times 10^{12} A/m^2$. The grid size was taken to be $ 4 \times 4 \times 0.6 nm^3$. Fig. \ref{fig:calib}(b) depicts the variation of the domain wall velocity with input charge current density. The velocity increases linearly with the current density and ultimately reaches a saturation velocity. The graphs are in good agreement with results illustrated in \cite{martinez2014current} for the same multilayer structure described in this section. 

\begin{figure}
\centering
\includegraphics[width =3.2in ]{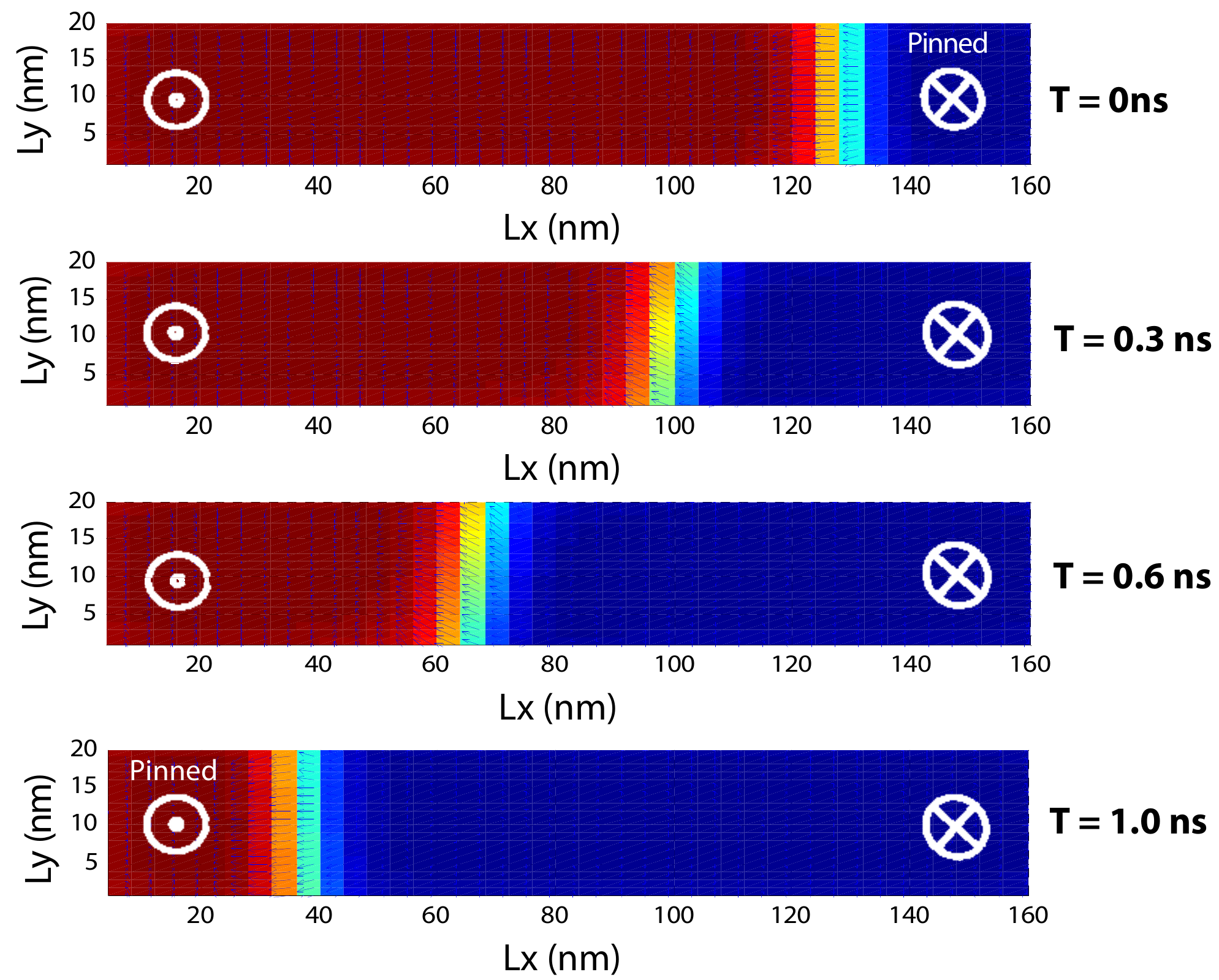}
\caption{Domain wall motion in the device due to programming current of $25 \mu A$ flowing through the HM underlayer for a duration of $1 ns$. The FM was taken to be $120nm$ in length surrounded by pinned layers of length $20nm$ on either side. The domain wall is displaced entirely from one edge of the FM to the other edge.}
\label{fig:dwm}
\end{figure}

It is worth noting here that for a given duration of the current through the heavy metal, the domain wall displacement is directly proportional to the magnitude of the current (considering input current range to be less than the saturation regime). The simulations were performed in MuMax3, a GPU accelerated micromagnetic simulation framework \cite{vansteenkiste2014design}. Fig. \ref{fig:dwm} shows the temporal motion of the DMI stabilized domain wall in the device due to a programming current flowing through the HM for a duration of $1ns$. For a device with ``free layer'' dimensions of $120nm \times 20nm$ , a maximum current of $\sim 25 \mu A$ is required to displace the domain wall from one edge of the FM to the other edge in a duration of $1ns$. 
\begin{figure}
\centering
\includegraphics[width = 3.3in ]{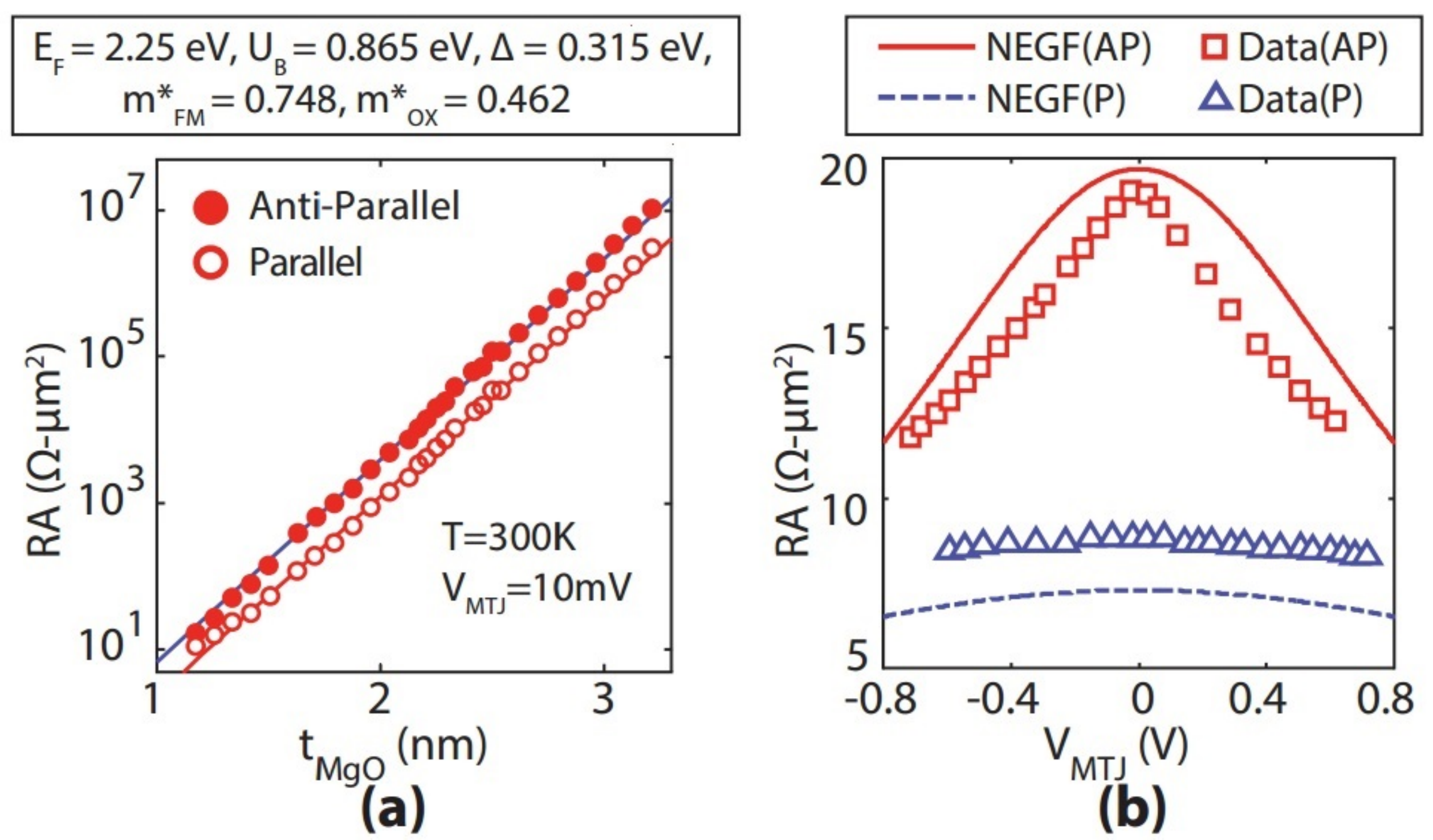}
\caption{The NEGF based transport simulation framework was calibrated to experimental results illustrated in \cite{yuasa2004giant,lin200945nm}. (a) Device resistance increases with increase in oxide thickness. (b) The AP MTJ resistance decreases with increase in the applied voltage across the MTJ. However, for sufficiently low values of applied voltage ($<100 mV$), the AP resistance variation is extremely small.}
\label{fig:calib_mtj}
\end{figure}

In order to simulate the variation of the MTJ resistance with domain wall position, Non-Equilibrium Green's Function (NEGF) based transport simulation framework \cite{fong2011knack} was utilized. The simulation framework was calibrated to experimental results illustrated in \cite{yuasa2004giant,lin200945nm}. For determining the MTJ resistance for a FM with a domain wall separating two oppositely polarized magnetized domains, the NEGF based simulator \cite{fong2011knack} was modified by considering the parallel connection of three MTJs. The magnetization direction of the FL of the three MTJs were considered parallel, anti-parallel and perpendicular (domain wall) to the pinned layer magnetization. The length of the first two MTJs was varied according to the position of the domain wall while the width of the third MTJ was taken to be equal to the domain wall width. Additionally, as shown in Fig.~\ref{fig:calib_mtj}, the resistance range of the device can be varied by varying the oxide thickness. 

\subsection{System Level Simulations}
The accuracy of the neural network over the image set was evaluated by varying the bit discretization level in the neurons and synapses. It was observed that there was insignificant degradation in accuracy with 15 (4 bit) intermediate levels and 3 (2-bit) intermediate levels between the maximum and minimum values of the synapse weights and neuron outputs respectively. Assuming that domain wall displacement over a distance of $10nm$ can be sensed and considering a domain wall width of $\sim 10nm$ (approximately), the length of the ``free layers" of the neurons and synapses were chosen to be $50nm$ and $170nm$ respectively.

Let us first discuss the simulation results for the spintronic neuron. Since the neuron is the computing element in the network, the device can be aggressively scaled down. Hence the area of the neuron ``free layer" was taken to be $50nm \times 20nm$, corresponding to an MTJ ``pinned layer" area of $30nm \times 20nm$. The critical current required to displace the domain wall from one edge of the ``free layer" to the other was observed to be $\sim 5 \mu A$ for a switching time of $2ns$. The synaptic crossbar array was split up into ``positive" and ``negative" arrays as mentioned in the previous section (corresponding to positive and negative synaptic weights). System level simulations yielded a maximum synaptic current of $\sim 50\mu A$ as input to the neuron. The resistance lying in the path of the input synaptic current was estimated to be $~\sim 140 \Omega$ (for the above mentioned device dimensions) resulting in a maximum voltage drop of $\sim 7mV$ across the spintronic neurons. 

Such ultra-low voltage operation of the spintronic neurons help in reducing the overall power consumption of the All-Spin neural network since the crossbar arrays can be operated at a much lower voltage. In our simulations, we considered a maximum input voltage of $100mV$ across each horizontal row of the crossbar array. The minimum resistance (corresponding to a synaptic weight ``1") in the crossbar array was evaluated to be $20K\Omega$ (to ensure the critical current requirement of $5\mu A$ to move the domain wall from one edge of the ``free layer" to the opposite edge). Based on the crossbar resistance values, the maximum value of $\gamma$ in the network was determined to be $\sim 0.07<<1$, thereby validating the assumption that the voltage drop across the spin-neurons is negligible. Further, it is also apparent from Fig. \ref{fig:calib_mtj} that variation in AP or P resistance of MTJ is extremely small for applied voltages $<100mV$. Hence, variation in synaptic conductance with applied voltage was considered to be insignificant. The maximum current flows through the synapse when $100mV$ is applied across the minimum synaptic resistance ($20K\Omega$), which is $\sim 5\mu A$. In order to ensure that the ``read" current flowing through the synapse does not cause any domain wall displacement, the width of the synapse was scaled up to $200nm$. For the above device dimensions, the heavy metal resistance lying in the path of the synaptic current was $\sim 50\Omega$ which is much lower than the range of synaptic resistances in the crossbar array.

\begin{figure}
\centering
\includegraphics[width =1.8in ]{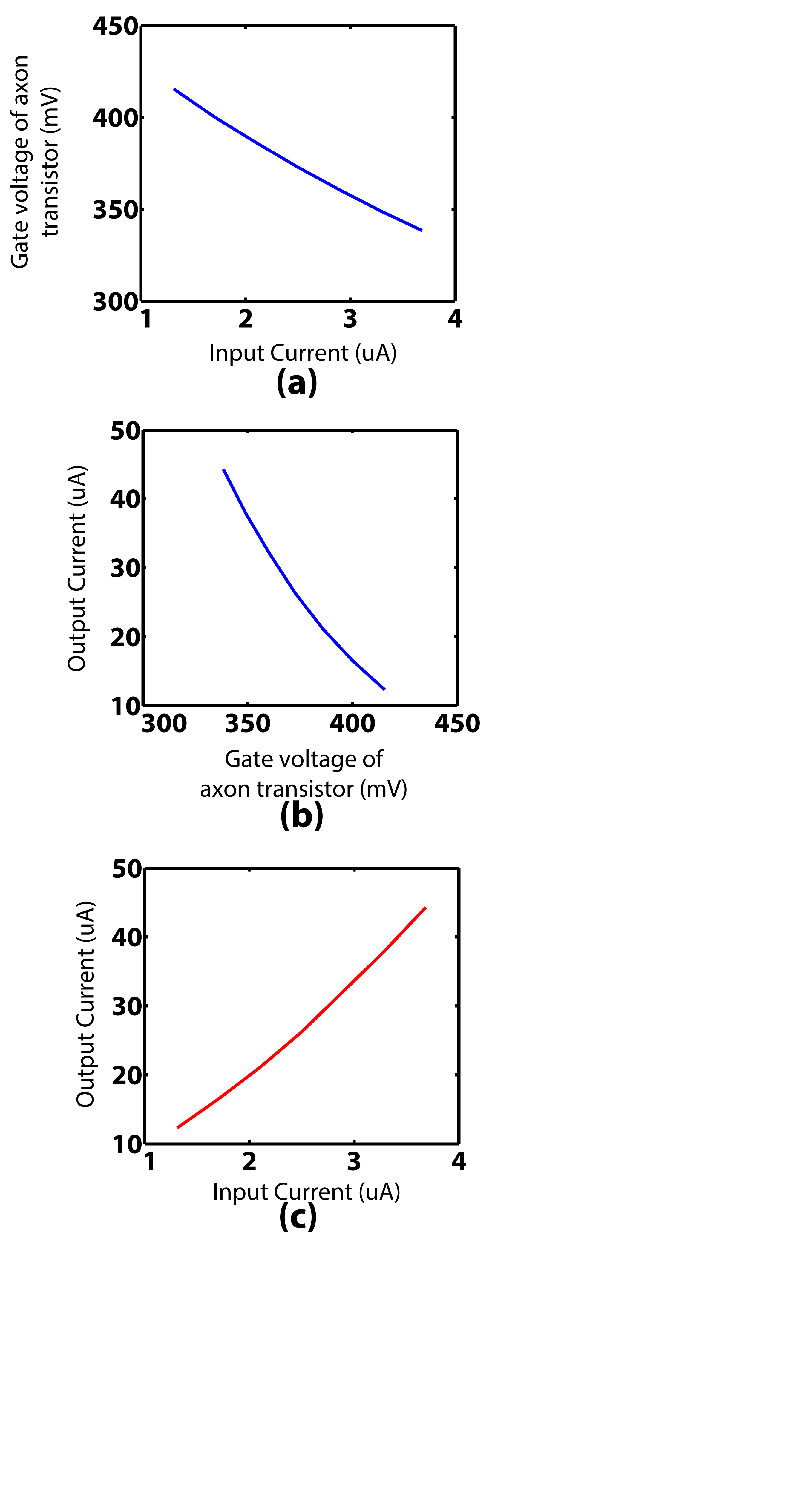}
\caption{(a) Gate voltage of axon transistor decreases with increase in magnitude of neuron input current, (b) Output current provided by axon transistor reduces with increase in the gate voltage, (c) Output current provided by the axon transistor increases almost linearly with the input current to the neuron. Hence, the neuron transfer function was taken to be linearly increasing with the input, ultimately saturating at a maximum value.}
\label{trans_fn}
\end{figure}
Fig. \ref{trans_fn} illustrates the variation of the output current provided by the axon transistor with input current provided to the neuron. As the magnitude of input current flowing through the heavy metal underlayer of the neuron increases, the gate voltage, $V_G$, of the axon transistor decreases as the pull-down resistance of the resistive divider network decreases. The supply voltage of the PMOS axon transistor was taken to be $650mV$. The supply voltage of the resistive divider network ($0.9V$) was optimized such that the corresponding swing in the gate voltage resulted in maximum swing of the output current. The maximum current to be supplied by the axon transistor was determined by the value of $G_{eq}$, such that the maximum voltage across the crossbar array was $\sim 100mV$. As shown in Fig. \ref{trans_fn}(c), the output current provided by the axon transistor increases almost linearly with the input current to the neuron.

Another important point of consideration is the degradation of classification accuracy due to device mismatches and variability in the spintronic devices. Although there can be variation in the programming of domain wall position in the spintronic neurons/synapses, deterministic domain wall motion has been observed in such magnetic multilayer structures by several research groups \cite{emori2013current,ryu2013chiral,bhowmik2014deterministic}. Further,
notches can be also utilized to pin the domain wall at specific locations along the length of the magnet to achieve necessary bit discretization \cite{lacour2004experimental}. Further, impact of resistance variation of the devices on the classification accuracy was evaluated by running 100 stochastic simulations of the network. Negligible degradation ($<4\%$) in classification accuracy was observed even with $20\%$ $3\sigma$ variation in the resistances of the spintronic devices. This can be attributed to the error resilient nature of such brain-inspired computing systems. Further, recent efforts in the implementation of on-chip or chip-in-the-loop learning may be utilized to alleviate such issues and implement variation immune neuromorphic systems \cite{prezioso2015training}.

Let us now discuss the energy consumption involved in the spintronic neuron. It has three components, namely, the ``write" power, the ``read" power and the ``reset" power. The All-Spin neural network was simulated over the entire set of 260 images and the magnitude of the average current flowing through each neuron was estimated to be $\sim 17.5 \mu A$ through the entire time window of $4ns$ ($2ns$ for the ``positive" crossbar array and $2ns$ for the ``negative" array). This results in an average ``write" energy consumption of $0.17fJ$ ($\sim I^{2}Rt$ energy consumption). For the ``read" circuit, the average current from the voltage supply ($0.9V$) was maintained to sufficiently low values ($\sim 80nA$) by an appropriate MTJ oxide thickness of $\sim 2nm$. Lower read current helps in ensuring that there is no domain wall displacement during the ``read" cycle and additionally helps in reducing the overall neuron power consumption. Since the ``read" circuit for the hidden layer has to provide the input current to the ``positive" and ``negative" crossbar arrays for the output layer, each for a duration of $2ns$, the overall ``read" energy consumption involved is $0.15fJ$ ($VIt$ energy consumption). Considering that a current of $\sim 5\mu A$ can ``reset" the neuron in a duration of $2ns$, the $I^{2}Rt$ ``reset" energy consumption is $\sim 0.007fJ$. As a result, the average overall energy consumption of the spintronic neuron is $\sim 0.32fJ$ which is almost two orders of magnitude lower in comparison to a corresponding analog ($\sim 700fJ$ \cite{sharad2013spin}) and digital ($\sim 832.6fJ$ \cite{ramasubramanian2014spindle}) CMOS neuron design in $45nm$ technology. Additionally, for a given range of synaptic resistances, the crossbar array can be operated at ultra-low voltages of $\sim 100mV$. In contrast, the crossbar arrays have to be operated at a much higher voltage $\sim 500mV (V_{dd}/2)$ for running analog CMOS neurons. This results in power savings by a factor $\sim 25 \times$ per synapse ($V^{2}/R$ power consumption) and thereby helps in reducing the overall power consumption of the neuromorphic system.

In conclusion, we have provided a vision for an All-Spin neural architecture where a single nanoelectronic device is able to mimic neuron and synapse functionalities in such systems. We provided extensive results for a standard image recognition problem based on an experimentally benchmarked device simulation framework to illustrate the functionality and energy-efficiency. Such All-Spin ANN architectures can potentially pave the way for ultra-low power deep learning neural systems.

\vspace{-2mm}

\bibliographystyle{IEEEtran}

\vfill
\begin{IEEEbiography}[{\includegraphics[width=1in,height=1.25in,clip,keepaspectratio]{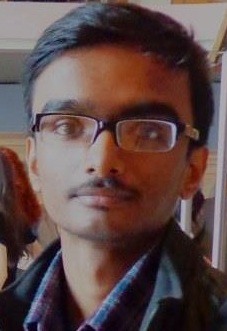}}]{Abhronil Sengupta}
received the B.E. degree from Jadavpur University, India in 2013. Currently he is pursuing Ph.D. degree in Electrical and Computer Engineering at Purdue University since Fall 2013. 

He is presently a Research Assistant to Prof. Kaushik Roy with the Nanoelectronics Research Laboratory, Purdue University. He worked as a summer intern and DAAD (German Academic Exchange Service) Fellow at the University of Hamburg, Germany in 2012. His primary research interests lie in low-power neuromorphic computing using spintronic and emerging devices.

Mr. Sengupta was the recipient of the Birck Fellowship from Purdue University in 2013 for his academic excellence.
\end{IEEEbiography}
\begin{IEEEbiography}[{\includegraphics[width=1in,height=1.25in,clip,keepaspectratio]{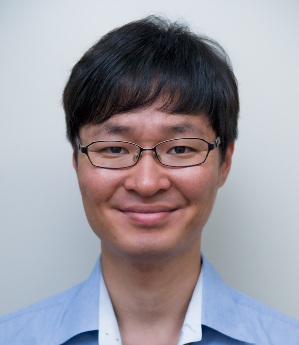}}]{Yong Shim}
received the B.S. and M.S. from Korea University in 2004 and 2006, respectively, all in Electrical Engineering. In 2006, he joined Samsung Electronics, Hwasung, Korea, where he was involved in designing circuits for memory interface. He is currently pursuing a Ph.D. degree as Williams Fellow and Graduate Research Assistant under the guidance of Prof. Kaushik Roy in the School of Electrical and Computer Engineering at Purdue University, West Lafayette, IN, USA. 

His research interests include emerging device based neuromorphic computing and approximate computing, as well as Complementary-Metal-Oxide-Semiconductor (CMOS) interface circuits for  spin-devices such as Spin-Torque Oscillator (STO). 

\end{IEEEbiography}
\begin{IEEEbiography}[{\includegraphics[width=1in,height=1.25in,clip,keepaspectratio]{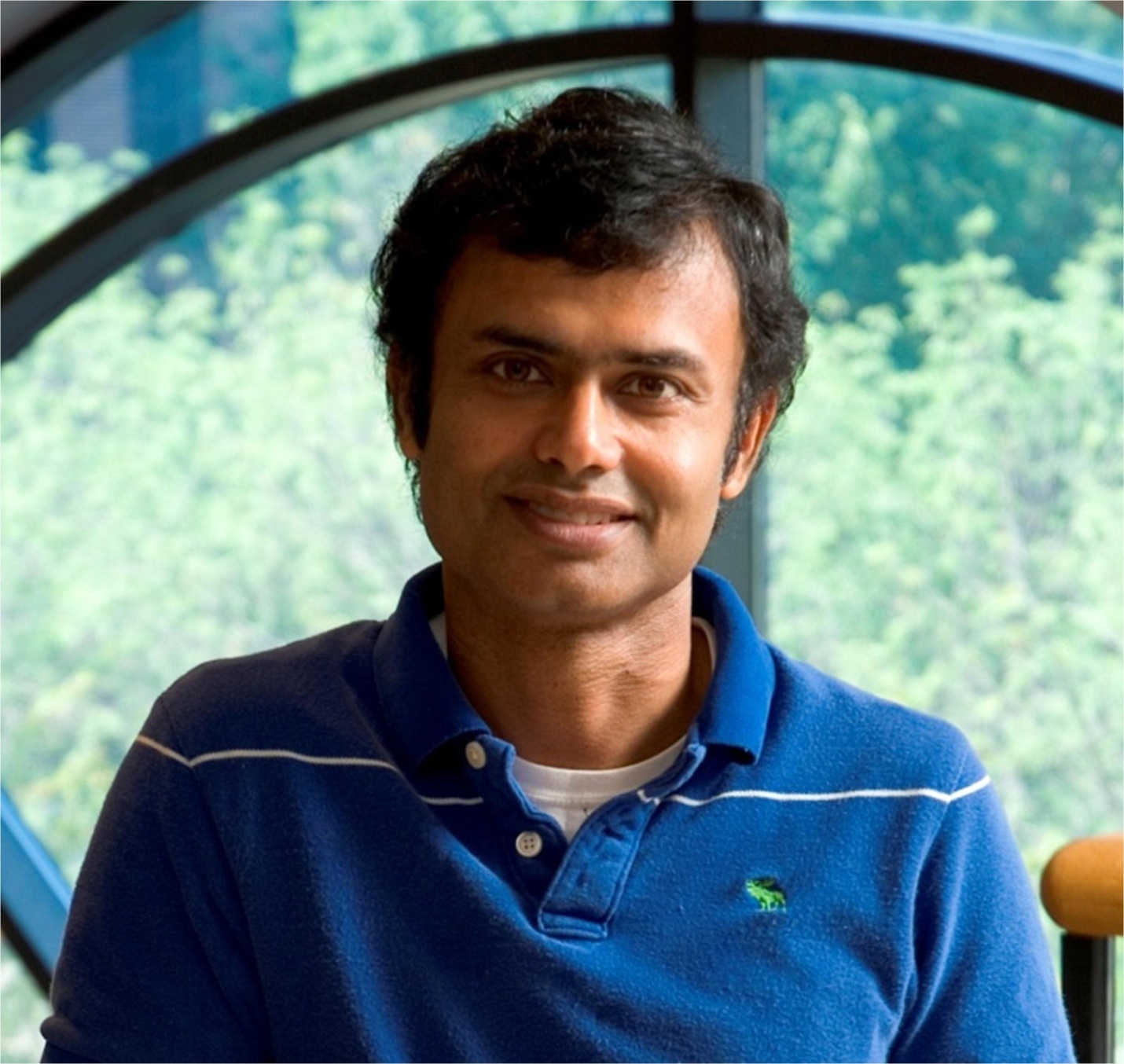}}]{Kaushik Roy}
received B.Tech. degree in electronics and electrical communications engineering from the Indian Institute of Technology, Kharagpur, India, and Ph.D. degree from the electrical and computer engineering department of the University of Illinois at Urbana-Champaign in 1990. He was with the Semiconductor Process and Design Center of Texas Instruments, Dallas, where he worked on FPGA architecture development and low-power circuit design. He joined the electrical and computer engineering faculty at Purdue University, West Lafayette, IN, in 1993, where he is currently Edward G. Tiedemann Jr. Distinguished Professor. His research interests include spintronics, device-circuit co-design for nano-scale Silicon and non-Silicon technologies, low-power electronics for portable computing and wireless communications, and new computing models enabled by emerging technologies. Dr. Roy has published more than 600 papers in refereed journals and conferences, holds 15 patents, supervised 65 PhD dissertations, and is co-author of two books on Low Power CMOS VLSI Design (John Wiley \& McGraw Hill). 

Dr. Roy received the National Science Foundation Career Development Award in 1995, IBM faculty partnership award, ATT/Lucent Foundation award, 2005 SRC Technical Excellence Award, SRC Inventors Award, Purdue College of Engineering Research Excellence Award, Humboldt Research Award in 2010, 2010 IEEE Circuits and Systems Society Technical Achievement Award, Distinguished Alumnus Award from Indian Institute of Technology (IIT), Kharagpur, Fulbright-Nehru Distinguished Chair, DoD National Security Science and Engineering Faculty Fellow (2014-2019), Semiconductor Research Corporation Aristotle award in 2015, and best paper awards at 1997 International Test Conference, IEEE 2000 International Symposium on Quality of IC Design, 2003 IEEE Latin American Test Workshop, 2003 IEEE Nano, 2004 IEEE International Conference on Computer Design, 2006 IEEE/ACM International Symposium on Low Power Electronics \& Design, and 2005 IEEE Circuits and system society Outstanding Young Author Award (Chris Kim), 2006 IEEE Transactions on VLSI Systems best paper award, 2012 ACM/IEEE International Symposium on Low Power Electronics and Design best paper award, 2013 IEEE Transactions on VLSI Best paper award. Dr. Roy was a Purdue University Faculty Scholar (1998-2003). He was a Research Visionary Board Member of Motorola Labs (2002) and held the M.K. Gandhi Distinguished Visiting faculty at Indian Institute of Technology (Bombay). He has been in the editorial board of IEEE Design and Test, IEEE Transactions on Circuits and Systems, IEEE Transactions on VLSI Systems, and IEEE Transactions on Electron Devices. He was Guest Editor for Special Issue on Low-Power VLSI in the IEEE Design and Test (1994) and IEEE Transactions on VLSI Systems (June 2000), IEE Proceedings -- Computers and Digital Techniques (July 2002), and IEEE Journal on Emerging and Selected Topics in Circuits and Systems (2011). Dr. Roy is a fellow of IEEE.
\end{IEEEbiography}

\end{document}